\documentclass{PoS}
\pdfoutput=1
\usepackage{graphicx}

\title{On the stability of particle dark matter}

\ShortTitle{On the stability of particle dark matter}

\author{\speaker{Thomas Hambye}\\
        Service de Physique Th\'eorique, Universit\'e Libre de Bruxelles, Bld du Triomphe, CP225, \\1050 Brussels, Belgium\\
        E-mail: \email{thambye@ulb.ac.be}}

\abstract{From the particle physics point of view, the most peculiar property of the dark matter particle is 
its stability on cosmological time scales. We briefly review the possible origins of this characteristic feature for candidates whose relic density results from the thermal freeze-out of their annihilation.
We emphasize that each stabilization mechanism implies an all specific phenomenology.
The models reviewed include supersymmetric and non-supersymmetric models where the stability is a consequence of grand-unification, models where stability is due to an unbroken gauge group and models where the DM stability is accidental. The latter possibility includes minimal dark matter, hidden vector dark matter and composite DM models.}

\FullConference{Identification of Dark Matter 2010-IDM2010\\
		July 26-30, 2010\\
		Montpellier France}

\begin{document}

\section{Introduction}

The dark matter (DM) particle is fantastically stable. Its decay lifetime has to be obviously larger than the age of the Universe, $\sim 10^{18}$~seconds. In most cases it has to be even larger than about $10^{26}$ seconds, in order that the decay doesn't cause fluxes of cosmic positrons, antiprotons and $\gamma$  larger than the observed ones. It turns out nevertheless that in most DM models this astonishing fact is not explained but rather assumed by hand, typically by assuming a discrete $Z_2$ symmetry. That, of course, does not mean at all that these models couldn't be correct, but certainly that their completeness is questionable. The origin of this stability is most probably preponderant in determining the structure of the DM interactions. Besides providing, for example, an explanation for the DM relic density, it would be highly desirable that DM models also provide a fundamental explanation for this stability. 
Here, by fundamental, we mean from a dynamical reason, resulting from a gauge symmetry and the particle field content of the model. A strong motivation to look for such stabilization mechanism is the situation of the Standard Model (SM). In the SM there is a series of stable particles whose stability  results in all cases from the gauge symmetries and particle content, rather than from an ad-hoc symmetry.
The photon is stable because it is the massless gauge boson of the exact electromagnetic $U(1)_{QED}$ gauge symmetry. The $e^-$ is stable because it is the lightest particle charged under this gauge group. The lightest neutrino is stable because of Lorentz invariance, since it is the lightest fermion. Finally the proton is stable because of conservation of baryon number, which is not imposed by hand, but results accidentally from the SM gauge symmetries and the gauge charges assigned to the SM particles. In the following we review briefly the various simple stabilization mechanisms which can be invoked. We stress that the stabilization mechanism has indeed a preponderant influence on the structure of the model and therefore on the associated phenomenology.
We will distinguish 2 kinds of mechanisms, the low energy ones, which similarly to the SM allow to understand the particle stability directly at low energy, and the high energy ones which rely on the existence of some explicit UV physics, typically on the existence of a high energy gauge group.
The latter possibility is less interesting phenomenologically because it relies on an assumption difficult to probe experimentally, but is certainly theoretically attractive too, especially if it is connected to grand-unification.

\section{The MSSM case}

In the minimal supersymmetric standard model, the lightest supersymmetric particle (LSP), in general assumed to be the lightest neutralino, is stable due to the R-parity symmetry.
R-parity has not been assumed originally to get a stable DM particle but to avoid proton decay. It is an attractive feature of the MSSM that DM stability results from a symmetry which has been assumed for an other reason. However, yet, R-parity is assumed totally by hand in the MSSM and is not explained. Moreover, to avoid proton decay, it would be sufficient to assume the conservation of baryon number or lepton number. This would allow terms which do break R-parity and hence would destabilize the LSP.

There is nevertheless a way to justify R-parity from a gauge principle. R-parity which is a symmetry distinguishing partners and super-partners, is connected to the superfield R-symmetry, under which all quark and leptons superfields are odd and the Higgs superfields are even. For any particle of spin $s$, $R_p=R_s \cdot (-1)^{2s}$. This symmetry prevents any R-parity breaking interactions. For any superfield of baryon number $B$ and lepton number $L$, $R_s$ turns out to be nothing but $R_s=(-1)^{3(B-L)}$. This means that $R$ symmetry is the discrete $Z_2$ subgroup of any gauge $U(1)_{B-L}$ group we could have on top of the SM gauge groups. Consequently it is is also a subgroup of the grand-unified $SO(10)$ group. Therefore $R$-symmetry could be the discrete remnant subgroup of these gauge groups. If all scalar fields, whose vacuum expectation values break these groups, are even under $B-L$, i.e.~have even $R_s$, $R$-symmetry will remain automatically unbroken \cite{mohap,Aulakh}, since they will leave all interactions even.
This is the well-known UV justification for R-symmetry. The phenomenology of these models is closely related to the fact that, in order that the mechanism works, all scalar representations breaking the various gauge symmetries have to be taken even. Natural in these theories are possible new states at the accessible TeV scale, $b$-$\tau$ unification, or scalar triplet seesaw as dominant origin of neutrino masses \cite{Aulakh,vissani}.

\section{Non-supersymmetric $U(1)_{B-L}$ justification of DM stability}

During a long time the above GUT justification of DM stability was somehow believed to be a privilege of supersymmetric theories. It is interesting to stress that this possible DM stability explanation is independent of supersymmetry. The same mechanism can be invoked in the non-supersymmetric context. This is a possibility which has been considered recently in Refs.~\cite{raidal1,raidal2,frigeriohambye}.
In the SM all quarks and fermions have odd $3(B-L)$ and in particular in $SO(10)$ an entire generation of them can be put in a single $16_\psi$ representation which is $3(B-L)$ odd. The Higgs fields on the other hand are even, and can be put in the $10_H$ representation of $SO(10)$, which is even. As a result if all fields breaking $U(1)_{B-L}$ and $SO(10)$ are $3(B-L)$ even (i.e.~among $10_H$, $45_H,\,54_H\,,120_H,\,126_H,\,210_H,\,...$), a $Z_2^{B-L}$ symmetry will remain unbroken. This leads to 2 possible classes of models for DM. Either DM is scalar in case to be stable it has to belong to an odd representation of $3(B-L)$, because all other scalar particle combinations it could couple to, to decay, are even. The $SO(10)$ odd representations it could belong to are $16_H,\,144_H,\,560_H,\,...$, as showed in Refs.~\cite{raidal1}-\cite{raidal4}. Or DM could be a fermion \cite{frigeriohambye}, if it belongs to an even representation, among $10_\psi,\,45_\psi,\,54_\psi\,,120_\psi,\,126_\psi,\,210_\psi,\,...$because all other fermion combination it could couple to are odd. This leads to two minimal models: either it belongs to a scalar $16_H$ representation, in case it is necessarily a combination of the electroweak singlet and the neutral component of the electroweak doublet in this representation. These are the only 2 neutral and colorless particles in this multiplet, and we know DM has to be neutral and colorless. Or it belongs to the fermion $45_\psi$ (or very similarly $54_\psi$) representation, in case it has to be the neutral component of the fermionic electroweak triplet in this 
representation. 

The phenomenology of the scalar candidate has been considered at length in Refs.~\cite{raidal1,raidal2,raidal3,raidal4}. The scalar DM particle stands within the range $\sim 70$~GeV-few~TeV, the lower bound coming from the LEP constraint on the search for charged Higgs particles. The direct detection cross section rate lies typically between $10^{-43}$~cm$^2$ and $10^{-47}$~cm$^2$, therefore in principle detectable in most cases, see Fig.~3 of \cite{raidal2}. The next-to-lightest dark scalar as well as the charged scalar DM partner in the doublet, are predicted to be long-lived in various regimes, providing experimental signatures in the form of displaced vertices with missing (DM) energy. Electroweak symmetry breaking can occur radiatively in this setup from the DM-Higgs couplings.

The fermion case \cite{frigeriohambye} is in our opinion interesting in many respects. First of all it is quite predictive. As the DM is made of the neutral component of the hyperchargeless fermion triplet, the only interactions it can have are the gauge interactions (unlike a scalar which can always have unknown quartic scalar interactions). It is therefore a pure "WIMP" candidate whose annihilation cross section (to gauge bosons or to a fermion pair) is completely fixed by the gauge interactions and DM mass. As a result the observed DM relic density $0.091< \Omega_{DM} h^2 < 0.129$ \cite{WMAP} can be obtained only for a fixed value of the DM mass which turns out to be $2.7\pm0.15$~TeV \cite{strumiamin1,hisano,strumiamin2}.
Recently, taking into account the effect the DM kinetic decoupling has on the Sommerfeld enhancement of the DM annihilation, the right relic density has also been found for $m_{DM}\simeq 4$~TeV, where a Sommerfeld resonance stands \cite{mohanty}.
Secondly since we justify the DM stability by grand-unification we must find a way in this model to get unification of the SM gauge couplings at the GUT scale, $M_{GUT}\sim10^{16}$~GeV.
It is remarkable that the adjunction to the SM of the fermion DM triplet pushes the unification scale (of the hypercharge and weak couplings) from $\sim 10^{12}$~GeV scale to the GUT scale (which is sufficiently high to avoid proton decay problems).
In other words if DM is a WIMP, which is the most straightforward possibility, we know it has to be around the electroweak scale or below. It is therefore fully motivated to believe, on the basis of the DM experimental fact, that there should be at least one extra particle on top of the SM ones around the electroweak scale. It turns out that, adding just this DM WIMP, one can have unification of the gauge couplings.\footnote{Of course, to achieve unification of the 3 gauge couplings, one also needs as usual to add colorful particles, in order to increase the value of the QCD coupling at the GUT scale. This is provided by the setup too, as the 45 (or 54) fermion representation contains a color octet which can do the job \cite{frigeriohambye}.} In other words this model succeeds in realizing in a particularly minimal way the features split-supersymmetry was mostly made for (i.e.~DM and gauge unification), without needing to assume any supersymmetry. Of course this model, as split supersymmetry, doesn't address the hierarchy problem. However to have an extra fermion around the electroweak scale doesn't appear as unnatural in the sense that it is protected by a $U(1)$ global chiral symmetry in the limit of vanishing
DM mass.

The direct detection rate of a $2.75$~ TeV fermion triplet has been determined in Ref.~\cite{strumiamin1}. It is of order $10^{-45}$ cm$^2$, hence in principle reachable. As for indirect detection, a promising possibility is to observe, in atmospheric Cherenkov experiment looking at the galactic center \cite{cherenkov}, monochromatic $\gamma$ lines from its annihilation. The corresponding rates for a 4~TeV candidate are given in Ref.~\cite{mohanty}, where the production of positrons has been found to be largely boosted by the Sommerfeld effect, as required to reproduce the $e^+$ excess observed by the Pamela satellite.

\section{Exact gauge symmetry setup}

In the following we will review 5 setups to justify the DM stability in a way which can be understood directly at low energy, i.e. without needing to assume a high energy gauge symmetry. Three of them, section 6, rely on the existence of an accidental symmetry. The fourth one, section 5, relies on the existence of a remnant symmetry of a gauge group broken at low energy. The fifth one, in this section, relies on the existence of an exact low energy gauge symmetry. 

In the SM the $e^-$ is stable because it is the lightest charged particle under the unbroken electromagnetic gauge group, $U(1)_{QED}$.
Similarly if there were, on top of the SM gauge groups, an extra unbroken $U(1)_{QED'}$,  the lightest charged particle under it would be stable. Actually it is impressive that the adjunction to the SM of the simplest gauge structure one could think of, that is to say QED for a single fermion, 
\begin{equation}
{\cal L} ={\cal L}_{SM} +  \bar{\psi} (i\!\not \hspace{-1.2mm}D -m_\psi) \psi 
%+\kappa F^Y_{\mu\nu} F^{\mu\nu}\,,
\label{qedprime}
\end{equation}
with $D_\mu=\partial_\mu-i e' A_\mu$, leads to a viable DM candidate! 
This QED' model, on top of the SM particles, contains an extra massless $\gamma'$  and an extra electron, $e'$ (singlet of the SM gauge groups), and nothing more! It has been studied in Ref.~\cite{kamion}, and in Refs.~\cite{feng2,feng} (in the context of a larger framework consisting of a copy of the MSSM in the hidden sector, with the DM in the form of a stau).
This structure also appears in the context of the mirror models, although in these models DM is in general dominated by the mirror baryons rather than by the mirror electron. Unless the QED' fine structure constant $\alpha'$ is tiny, DM is in thermal equilibrium with the $\gamma'$.
The DM sector, however, as considered in Refs.~\cite{kamion,feng2,feng}, is not in thermal equilibrium at all with the SM visible one, since no connector between both sectors are basically considered.
The relic density is determined by the usual freeze out of its pure hidden sector annihilation cross section $e' \bar{e}' \rightarrow \gamma' \gamma'$. It depends only on 3 parameters:  $m_{e'}$, $\alpha'$  and the temperature of the hidden sector relative to the SM 
one, $\xi \equiv T'/T$. The S-wave dominated cross section is $\sigma_{annih}\cdot v \simeq \pi \alpha'^2/(2 m^2_{DM})$.
% which gives a relic density
Analytically, for the freeze-out temperature $T_f$ we obtain:
\begin{eqnarray}
x_f&=&\xi \cdot ln[0.038 \cdot\xi^{5/2}\sigma_{annih}v M_{pl}m_{DM}
%c(c+2)
\frac{g_x}{\sqrt{g_{*}}}]\nonumber\\
&&-\xi \frac{1}{2}\cdot ln\{\xi \cdot ln[0.038 \cdot \xi^{5/2}\sigma_{annih}v M_{pl}m_{DM} 
%c(c+2)
\frac{g_x}{\sqrt{g_{*}}}]\}
\label{xf}
\end{eqnarray}
with $x_f=m_{DM}/T_{f}$ which gives
\begin{equation}
\Omega_{DM} h^2=\frac{1.07\times 10^9 x_f \cdot GeV^{-1}}{({g_{*S}}/\sqrt{g_{*}})M_{pl}\sigma_{annih}v}.
\label{omegaeprime}
\end{equation}
\begin{figure}[t]
%\begin{tabular}{c}
\begin{center}
\vglue -.1cm
\includegraphics[width=7.0cm]{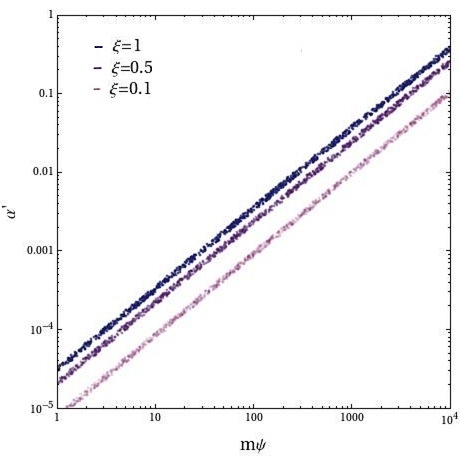}
%%%%\end{tabular}
\end{center}
\vglue -.8cm
\caption{Values of $\alpha'$ needed to get the observed relic density, as a function of $m_{DM}$ (in GeV), for $m_h=120$ GeV and $\xi$=0.1, 0.5, 1.}
 \label{fermion}
 \vspace{-3mm}
\end{figure}

Here it has been assumed that the present value of $\xi^3$ is small, so that $g_{*S}$(today)$\approx 3.91$ (which implicitly means that the Hubble expansion rate $H$ is mostly determined by the visible sector particle species). In Fig.~\ref{fermion} we display the values of $m_{e'}$ and
  $\alpha'$ which lead to a relic density within the WMAP range. 
They can be understood in the following way.  
Given the non-relativistic Boltzmann suppressed behavior of the DM equilibrium number density around freeze-out,  $n_{DM}(T')\propto (T')^{3/2}e^{-m_{DM}/T'}$, 
the value of $m_{DM}/T'_f=x_f/\xi$ doesn't change much with $\xi$, even though the smaller $\xi$, the larger the Hubble constant at $T'\simeq T'_f$ (when $\Gamma_{annih}\simeq H$). The dependence is only logarithmic. But if one considers a smaller $\xi$ the number of DM particles left at freeze-out increases as $H\sim \xi^{-2}$ (since $\Gamma_{annih}(T'_f)\simeq \sigma_{annih} v \,n_{DM}(T'_f)$). On the other hand the entropy at freeze-out, which is proportional to $T_f^3$, increases as $\xi^{-3}$. Therefore $\Omega_{DM}=m_{DM} \,n_{DM}^{Today}/\rho_{Today}\simeq m_{DM} (n^{DM}(T'_f)/s(T'_f)).(s_{Today}/\rho_{Today})$ is proportional to $\xi$, see Eq.~(\ref{omegaeprime}). For smaller $\xi$ this must be compensated by taking a smaller $\sigma_{annih}$, as shown in  Fig.~1, so that the freeze-out occurs earlier. Note that these results agree
with the one of Ref.~\cite{feng2} up to the power 5/2 instead of 3/2 in Eq.~(\ref{xf}) (which is freeze-out prescription dependent and has a moderate numerical effect).\footnote{These results have been obtained in collaboration with Xiaoyong Chu, to appear in a longer publication.}

The QED' model is a typical example of model where the mechanism at the origin of DM stability  implies an all specific phenomenology. The presence of an extra unbroken gauge group implies extra radiation in the Universe, i.e.~the presence of a long range force which acts on the charged DM particles.
This has an all series of cosmological and astrophysical implications \cite{kamion,feng2,feng}, such as a)
$e'$--$\bar{e}'$ boundstate formation which can change the DM annihilation rate at various epochs, b) delaying of the DM kinetic decoupling from $\gamma' e'\rightarrow \gamma' e'$ Compton scattering which can delay the formation of small scale structure, c) DM self interactions induced by $e' e' \rightarrow e' e'$ Rutherford scattering which may affect the dynamics of the Bullet Cluster as well as the morphology of DM galactic haloes. The main constraint turns out to be the latter: more collisions affects the ellipticity of the DM halo, tending to the formation of a more isothermal profile, that is to say to the formation of a core, a region of constant DM density around the galactic center. 
It gives an upper bound on $\alpha'$ which decreases if $m_{DM}$ decreases. The upper bound is $\alpha'\lesssim 10^{-8},\,10^{-5},\,10^{-2}$ for $m_{DM}=100\,\,\hbox{MeV},\,10\,\,\hbox{GeV},\,1\,\,\hbox{TeV}$ respectively. Combined with the relic density constraint above it gives the upper bound $\xi < 10^{-5},\,10^{-3},\,10^{-1}$ respectively.
In comparison, the BBN constraint on extra relativistic degrees of freedom at $T_{BBN}\sim 1$~MeV is very mild. If in the hidden sector all particles have a mass exceeding this temperature, except the $\gamma'$, the constraint is $\xi \lesssim 0.9$.

Given these small values of $\xi$, the connectors between this $QED'$ structure and the SM visible sectors must be small, otherwise both sectors would thermalize, giving~$\xi\simeq1$. For the fermionic case of Eq.~(\ref{qedprime}) the only possible (i.e.~gauge invariant) term is a kinetic mixing term ${\cal L} \owns\kappa F^Y_{\mu\nu} F^{\mu\nu}$, with $F_{\mu\nu}^Y$ the hypercharge gauge field. One can determine that $\kappa$ has to be smaller than $10^{-9}$--$10^{-11}$ to avoid a too large transfer of energy from the visible to the hidden sector (compared to the ellipticity bound on $\xi$ above).
This unfortunately prevents us to have access to the hidden sector and DM at colliders.

In the case of a scalar DM candidate, i.e.~scalar electrodynamics for a single hidden sector $\phi'$ field, the phenomenology is very similar, apart for one important difference: this DM scalar can directly interact with the SM through the Higgs portal $\lambda_m \phi' \phi'^\dagger H^\dagger H$ interaction. This interaction has nevertheless to be small for the same reason as for $\kappa$. Unfortunately, simplicity doesn't always mean testability at colliders.

\section{Remnant symmetry of a gauge group broken at low energy}

In the above one has seen that DM stability could result from the existence of a remnant global symmetry subgroup of a high energy gauge group. This requires that the gauge group is broken by scalar fields which are even under the remnant group. This, actually, could hold at low energy too, either by considering that $U(1)_{B-L}$ is broken at low energy, or independently of $B-L$. 
The simplest possibility in the latter case is to consider an extra $U(1)'$ gauge group with 2 particles charged under it, a scalar $\phi'$, which breaks it spontaneously, and a fermion $\psi'$. The relevant lagrangian is
\begin{eqnarray}
{\cal L}={\cal L}_{SM} + \frac{1}{4} F_{\mu \nu}F^{\mu \nu}+\bar{\psi'} (i\!\not \hspace{-1.2mm}D -m_{\psi'}) \psi' +
D_\mu \phi' D^\mu \phi'^\dagger \nonumber \\
-\mu^2_{\phi'} \phi' {\phi'}^\dagger -\lambda_{\phi'} (\phi' {\phi'}^\dagger)^2 -\lambda_m \phi' {\phi'}^\dagger H H^\dagger\,
+\kappa F^Y_{\mu\nu} F^{\mu\nu} \,,
\end{eqnarray}
with $H$ the SM Higgs doublet and with $\mu^2_{\phi'} <0$, so that $\phi'$ acquires a vev, $\langle \phi' \rangle\equiv v_\phi$. After symmetry breaking the model contains the DM fermion, a real scalar, $\eta$, and a massive $Z'$ gauge boson.
Since $\phi'$ has a vev, $\eta$ is obviously not stable but $\psi'$ remains stable even though $U(1)'$ is broken, simply because all interactions involve it in pairs.
Such a model has been considered in Ref.~\cite{Pospelov:2007mp}, in the so-called "secluded" DM framework. Since the gauge group is broken, the $U(1)'$ gauge boson is massive. Therefore all the constraints due to extra radiation in the QED' model above are not relevant anymore.
The interactions connecting the hidden and visible sectors, kinetic mixing and Higgs portal, can be much larger than for the unbroken QED' model above. For instance $\kappa$ can be as large as few $10^{-3}$. The relic density can proceed from the purely hidden sector process involving the $Z'$, in case it is "secluded" or, if the connectors are large, can result from the annihilation induced by these connectors. Interestingly, through the exchange of a $Z'$, the kinetic mixing interaction $\kappa$ induces an interaction between a DM pair and a SM fermion pair, $\bar{\psi}' \psi-> Z'->\gamma,Z\rightarrow \bar{f}_{SM} f_{SM}$.  If the DM mass is of order 10 GeV, one can get \cite{mambrini} from this single diagram both the observed relic density and a direct detection rate of order the ones indicated by the Dama and Cogent experiment \cite{Dama,Cogent}. This requires nevertheless a rather precise correlation between $m_{DM}$ and $m_{Z'}$: the mass of the $Z'$ must be within a range close to twice the DM mass value, in order that the DM annihilation is largely enhanced by a $Z'$ resonance. 

\section{Accidental stability setups}

As said above there exists already in the SM an example of particle which is accidentally stable, the proton. There is no reason why DM could not be stable in a similar way. Actually one would expect in this case an interesting property to hold: decay induced fluxes of cosmic rays of order the ones observed.
If the stability is accidental, there is in particular no reason why DM should be absolutely stable, unlike in the DM models considered above. Even if the pure low energy renormalizable interactions with the SM particles
do not allow DM to decay, in this case there is no 
symmetry protecting DM from being destabilized by 
an interaction involving a higher energy particle beyond the SM.
On the contrary, if from the exchange of such a heavy particle, a dimension 5 interaction (suppressed by one power of the heavy particle scale $\Lambda$), is induced, it will induce a far too short lifetime. The expected DM lifetime in this case is $\tau_{DM}\sim 8 \pi \Lambda^2/m^3$ where $m$ can only be a combination of the DM mass and SM particle masses. Considering a DM mass below few TeV leads to a lifetime shorter than the age of the Universe, $\tau_U$, unless $\Lambda \gtrsim 10^{25}$~GeV, i.e. far above the Planck scale. 
An even larger scale $\sim10^{29}$~GeV is needed, if one demands that the lifetime is larger than $\sim 10^{26}$~sec, what is mandatory for most models in order that the DM decay does not induce a flux of cosmic rays larger than the ones observed. One therefore must make sure that no dimension five operator between the DM candidate field and SM fields is allowed by the SM (and hidden sector) gauge symmetries.
But if, on the contrary, they allow dimension 6 operators it turns out that the lifetime, $\tau_{DM}\sim 8 \pi \Lambda^4/m^5$, is of order $10^{26}$~sec if $\Lambda\simeq 10^{14}$~GeV, which is close to the GUT scale! In other words an accidentally stable DM candidate which can be destabilized by a dimension 6 GUT scale induced interaction leads to a flux of cosmic rays of the order of the ones observed, and therefore potentially to a rich phenomenology.

\subsection{Minimal Dark Matter}

Probably the simplest setup which leads to an accidentally stable DM candidate is the Minimal DM one \cite{strumiamin1,strumiamin2,strumiamin3}. This framework doesn't require any extra gauge groups on top of the SM ones or the assumption of a global group. It requires, instead, high $SU(2)_L$ multiplets. 
It stems from the fact that if one considers a large enough fermion or scalar $SU(2)_L$ multiplet, $SU(2)_L$ invariance prevents it to have any "fast" renormalizable or dimension 5 interactions. For instance, to have a renormalizable interaction involving a fermion $SU(2)_L$ quadruplet, one needs it to couple to a quadruplet of dimension 5/2 made of SM fields. This can only involve a fermion, which in the SM belongs at most to a doublet, and the SM Higgs doublet scalar. From 2 doublets one gets at most a triplet.
Therefore a fermion quadruplet or higher has no renormalizable interactions with SM fields, but still can have a dimension 5 interaction (with one lepton doublet and 2 Higgs doublets).
But a fermion quintuplet or more cannot have a dimension 5 interaction either.
Similarly, for a scalar candidate, a sextuplet or higher cannot have any of these interactions.

To this constraint come 2 other ones. First DM must be neutral, hence the multiplet must have a neutral component, i.e.~an integer hypercharge. Second this neutral component cannot have any tree level interactions to the $Z$ boson, otherwise the direct detection rate driven by $Z$ exchange would be much larger than allowed experimentally. In other words the electric charge $Q$ must vanishes as well as the hypercharge $Y$ and the weak isospin $T_3$, which is possible only for an odd multiplet.
The good candidates are therefore a fermion quintuplet, septuplet,... or a scalar septuplet, nonuplet,...
Why there would be beyond the SM so large multiplets, with no other TeV scale smaller multiplets (which would allow to write again fast destabilizing interactions), is not clear but this is certainly a possibility.

We will not review here the phenomenology of such a candidate. It has been summarized in Ref.~\cite{strumiamin3}. Some of its main features are that a) the fermion quintuplet must have a 9.6~TeV mass in order to satisfy the relic density constraint, b) the direct detection rate is of order $10^{-44}$~cm$^2$, c) the indirect detection rate is boosted by two orders of magnitude by the Sommerfeld effect, so that the Pamela $e^+$ excess can be reproduced if one assumes an extra astrophysical boost of order 50 (but giving in this case a flux of $e^++e^-$ exceeding the Hess experiment value at energies above $\sim 2$~TeV), d) given the high DM mass this doesn't come with a too large flux of $\bar{p}$ (because the excess of $\bar{p}$ it gives is  above 100~GeV), and e) the expected flux of $\gamma$ produced in the galactic center tends to exceed the experimental limits in the case of a NFW halo profile, but a slightly more isothermal profile is fine.

\subsection{Hidden vector DM}

Another setup based on an accidental stabilization is the hidden vector DM model \cite{TH,HT,AHIW}. This model  in many respects is different from the other models above. First, this model shows that a non-abelian gauge boson could be a DM particle candidate. Second, its stability results from the automatic presence of an accidental symmetry which is non-abelian, unlike all models above which are based on a discrete or global $U(1)$ symmetry. Third, DM can be destabilized only by dimension 6 or higher operators. No need to impose that there are no dimension 5 operators. Fourth, related to the fact that DM has spin-1, the dimension 6 operators generically lead to intense monochromatic $\gamma$ ray lines directly at tree level, rather than at the usual suppressed one-loop level. As well known the observation of such lines would be a real smoking gun for DM, because there is no expected relevant astrophysical background for such a signal.

This model works as follows. Consider a non abelian gauge structure on top of the SM ones. The simplest example is to consider a $SU(2)$ one. Consider also a scalar multiplet of it, which breaks $SU(2)$ completely, in such a way that all 3 gauge bosons becomes massive (no extra radiation in the Universe). The most general Lagrangian for such a structure is
\begin{equation}
{\cal L}= {\cal L}^{SM} -\frac{1}{4} {F'}^{\mu\nu} \cdot F'_{\mu \nu}
+(D_\mu \phi)^\dagger (D^\mu \phi) -\lambda_m \phi^\dagger \phi H^\dagger H-\mu^2_\phi \phi^\dagger \phi -\lambda_\phi (\phi^\dagger \phi)^2 \,,
\label{inputlagr}
\end{equation}
with $D^\mu \phi=\partial^\mu \phi - i\frac{g_\phi}{2} \tau \cdot A'^\mu$.
% and the Higgs potential of the SM we define as: ${\cal L}^{SM} \owns -\mu^2 H^\dagger H -\lambda (H^\dagger H)^2$ with $H=(H^+,H^0)$. \\
It contains a pure gauge term with gauge coupling $g_\phi$, a kinetic term and potential for the scalar and a Higgs portal interaction. The latter is not forbidden by any of the visible or hidden sector gauge symmetry. 
After $\phi$ gets a vev, $v_\phi$, the 3 massive gauge bosons have a mass $m_V=g_\phi v_\phi/2$ and a single real scalar Higgs boson, $\eta$, remains. 
The all model can be parametrized in terms of 4 parameters: $g_\phi$, $m_V$, $m_\eta$ and $\lambda_m$.
Looking closer at this structure, which is similar to the scalar-gauge SM one (but without extra $U(1)$, i.e.~no Weinberg angle), what one remarks is that, even if the gauge group has been broken completely, a non-abelian global symmetry remains accidentally: a custodial $SO(3)$ group under which the 3 massive gauge bosons form a triplet vector and the $\eta$ a singlet. Since this symmetry is exact, any non singlet particle cannot decay into singlets, hence the massive gauge bosons are stable. Related to the custodial symmetry, the 3 gauge bosons have the same mass.
The custodial symmetry of the pure gauge-scalar hidden sector is not broken either by the Higgs portal interaction, i.e.~it can couple to the SM without being destabilized.

The massive vectors in this model can fulfill the various DM constraints. Their relic density results from the thermal freeze-out of their annihilation. We can distinguish 2 regimes depending on the size of the Higgs portal interaction. If $\lambda_m$ is small the relic density results from the pure hidden sector annihilation, $V_i V_i \rightarrow  \eta \eta$ and $V_i V_j \rightarrow V_k \eta$. Note that the last annihilation process is a 2 to 1 DM process, impossible in model based on a usual $Z_2$ symmetry.
It is induced by the non-abelian gauge trilinear term.
If $\lambda_m$ is instead large, typically larger than $10^{-3}$, extra annihilation to SM particles can be important or even dominates. Part of these channels comes from the fact that $\lambda_m$ induces a mixing of the $\eta$ boson with the SM Higgs boson $h$.
Values of $m_\eta$ and $m_V$ leading to the right relic density in both regimes cover wide ranges of values. 
Direct detection rates saturating the present experimental bounds as well as signatures at LHC can be easily obtained in the large Higgs portal regime. For more details see Refs.~\cite{TH,AHIW}.
Here instead we would like to discuss more the monochromatic $\gamma$ flux it is expected to lead to.

Since the custodial symmetry is accidental it could be broken, as explained above, by interactions induced from the exchange of higher energy fields. These interactions, which involve only low 
energy particles as external particles, have obviously to respect the gauge symmetries of both sectors.  Nicely there exist no dimension 5 operators of this kind, but 4 dimension 6 ones which break the custodial symmetry:
\begin{eqnarray}
\label{eqn:opA}
&{\rm (A)}&~~~\frac{1}{\Lambda^2}\  {D}_{\mu}\phi^{\dagger}\phi\ {D}_{\mu}H^{\dagger}H \,,\\
\label{eqn:opB}
&{\rm (B)}&~~~\frac{1}{\Lambda^2}\  {D}_{\mu}\phi^{\dagger}\phi\  H^{\dagger}{D}_{\mu}H \,,\\
\label{eqn:opC}
&{\rm (C)}&~~~\frac{1}{\Lambda^2}\  {D}_{\mu}\phi^{\dagger}{D}_{\nu}\phi\ F^{\mu\nu Y} \,,\\
\label{eqn:opD}
&{\rm (D)}&~~~\frac{1}{\Lambda^2}\ \phi^{\dagger} F^a_{\mu\nu}\frac{\tau^a}{2}\phi F^{\mu\nu Y}\,.
\end{eqnarray}
Associated to the spin 1 character of DM, all of them turn out to generically induce, with large branching ratios, 2 body decays involving a photon in the final state (from $F^{\mu\nu Y}$ or from the covariant derivatives of the SM Higgs doublet). As explained above, if the underlying high physics scale is of order the GUT scale one expects fluxes of cosmic rays of the order of the present sensitivities.
This is shown in Fig.~2 which, together with the flux of other cosmic rays produced, gives an example of monochromatic lines one can obtain assuming such a scale.
Note that, since these lines proceed through a decay, their magnitude is proportional to the number of DM particles $n_{DM}$, rather than to $n_{DM}^2$ as for an annihilation. As a result one doesn't need to look at the "involved" galactic center to expect an observable monochromatic line. The flux given in Fig.~2 is the extragalactic flux. Such monochromatic lines could be observed by the Fermi satellite if it increases its statistics, or could extend its analysis towards higher energies. 

Finally, about this model, note that if, instead of being broken by the vacuum expectation value of the scalar field $\phi'$, the $SU(2)_{HS}$ gauge structure of Eq.~(\ref{inputlagr}) confines at low energy (i.e.~if there is no $\phi'$ vev or if its value would be smaller than the confinement scale) the model leads to accidentally stable spin 1 states which are good DM candidates too \cite{HT}. In this case the stable vectors turn out to be boundstates of the scalar doublet and gauge bosons, of the type $\phi^\dagger D_\mu \phi$. They are stable because they also form a triplet of the custodial symmetry, and custodial symmetry is not broken by the confinement. Given the strongly interacting interactions it involves, DM particle is expected to stand in the multi TeV range, to satisfy the relic density constraint. 
\begin{figure}[t]
%\begin{minipage}[t]{0.5\textwidth}
  \centering
 { \hbox{  \includegraphics[width=0.9\columnwidth]{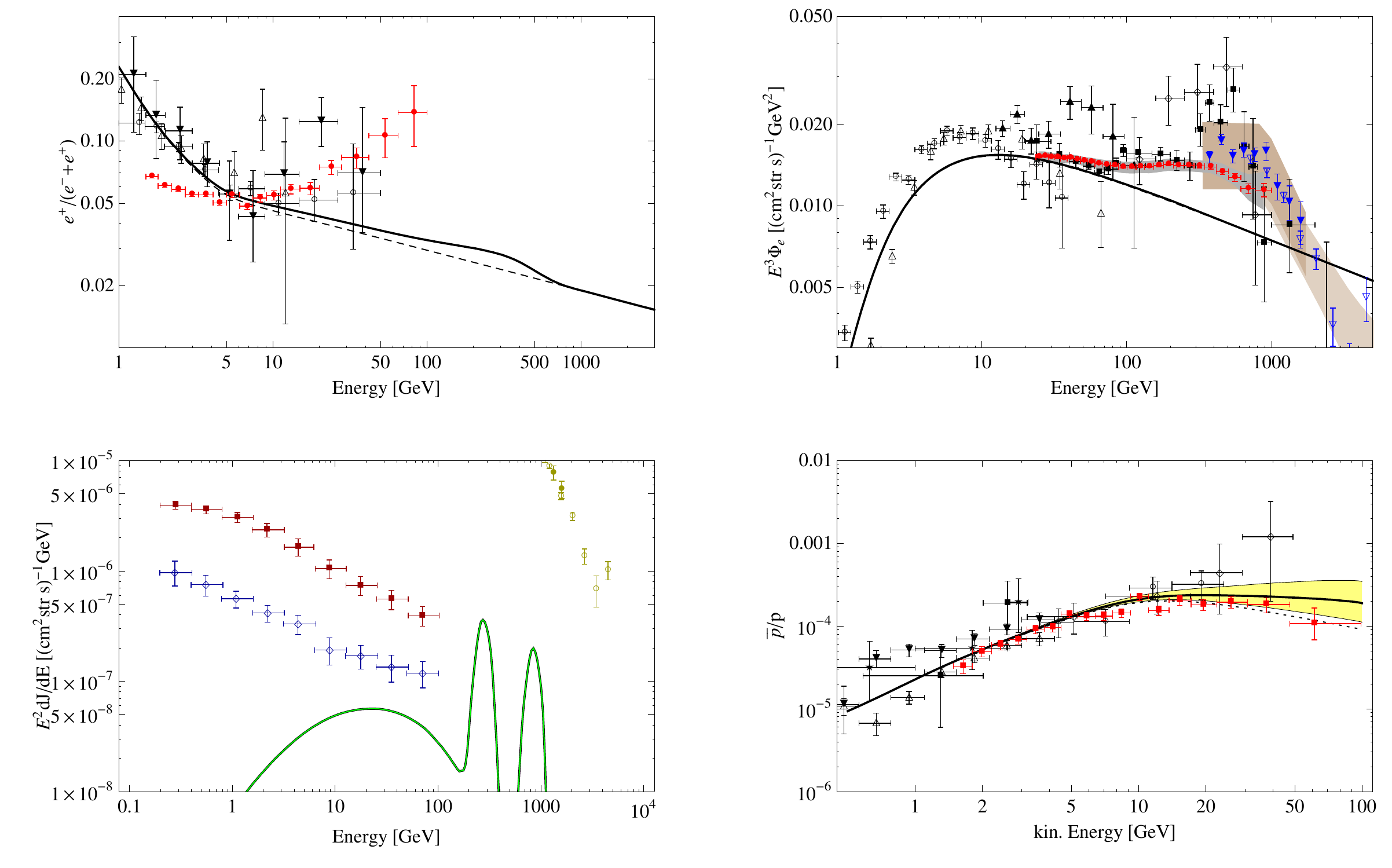}}}
  \centering
%\end{minipage} 
  \vspace{-0.2cm}
  \caption{Predictions \cite{AHIW} for $m_V=1550$~GeV,  $g_\phi=2.1$, $v_\phi=1457$~GeV, $m_\eta=1243$~GeV, $m_h=153$~GeV, and for the $\eta-h$ mixing, $\sin \beta=0.25$, considering as an example only operator C, with $\tau_{DM}=1.6\times10^{27}$~s ($\Lambda = 1.2\times10^{16}$~GeV). The
  \textit{upper panels} show the positron fraction (\textit{left}) and the
  total electron + positron flux (\textit{right}) compared with experimental
  data. \textit{Dashed} lines show the adopted astrophysical background,
  \textit{solid} lines are background + dark matter signal (which overlap the
  background in this plot). The \textit{lower left panel} shows the gamma-ray
  signal from dark matter decay, whereas the \textit{lower right panel} shows
  the $\bar{p}/p$-ratio: background (\textit{dashed line}) and overall flux
  (\textit{solid lines}, again identical with background).}
  \label{fig:Cf}
\end{figure}

\subsection{Weakly interacting stable pions}

Another possibility of accidentally stable DM candidate arises in the context of strongly interacting theories, as the last case discussed above, but in a different way. 
DM in this case are scalar boundstates made of fermions, similar to pions in the SM \cite{baihill}. Such a structure can arise if on top of the SM there is a new confining structure, a "QCD'" sector, under which there are new fermions which are vectorlike, i.e.~whose chiral components couple in the same way under the new strong interactions (as quarks for QCD) as well as under weak interactions (unlike quarks in the SM)
\begin{equation}
\mathcal{L}=\mathcal{L}_{SM}-\frac{1}{4} F_{\mu \nu}^a F^{\mu \nu a}+ \bar{\psi}_i ( i\partial\hspace{-2.1mm}/ \hspace{0.5mm}+ g_s A^b\hspace{-3.5mm}/ \hspace{2mm} t^b+g_W W^a\hspace{-4.3mm}/ \hspace{2mm} J^a) \psi_i 
\end{equation}
where $\psi_i=1,...,n$ are the new fermions, $t^a$ are the generators of the new strong group and $J^a$ are the SM generators of $SU(2)_L$. 
As in QCD, after the associated chiral symmetry gets dynamically broken, one obtains a hadronic mass spectrum whose lightest states are the chiral symmetry Goldstone bosons, the pions, "$\pi'$". As QCD too, the QCD' interactions respect isospin and charge conjugation, therefore respect the G-parity. G-parity conservation in QCD explains why a $\rho$ meson which has an even G-parity can decay strongly to 2 $\pi$, but not to 3 pions, since a pion has odd G-parity (and the contrary for the $\omega$ meson which has odd G-parity).
In the SM G-parity is nevertheless broken by the electroweak interactions due to the chiral structure of quarks under them, so that the lightest G-odd states, i.e.~the pions, are not stable. Here on the contrary since these new strongly interacting fermions are assumed to have the same electroweak interactions for left and right-handed components, G-parity is not broken and the $\pi'$ are stable.
The lightest $\pi'$ cannot annihilate strongly, since it is the lightest hadron, but do annihilate weakly. Therefore it is a real WIMP and has the right relic density if its mass is of order TeV.
As a result such a model can be probed at the LHC. For the direct detection the cross section is of order $10^{-45}$~cm$^2$, as for typical $SU(2)_L$ multiplets, see section 6.1.

Since G-parity is a purely accidental symmetry these pions could in principle decay from the exchange of heavy states. A problem one encounters in this framework is that, having a vectorial structure to guaranty the conservation of G-parity, one can write down G-parity violating dimension 5 operators destabilizing the pions, of the form $F^Y_{\mu \nu} \bar{\psi} \sigma^{\mu \nu} \psi/\Lambda$ and $H^\dagger J^a H \bar{\psi} J^a \psi/\Lambda$. The only way to avoid them appears to assume an extra global symmetry, such as a Peccei-Quinn symmetry. This weakens the naturalness of the stability of such candidates, but is certainly a possibility too.

Before concluding, note that another accidentally stable DM candidate arises in the context of the mirror models \cite{mirror}, in the form of mirror nuclei, protected by the conservation of mirror 
baryon number. Note also that
in the above we focus our attention on dynamical ways to justify the DM stability, because the gauge symmetry principle constitutes the most established and robust SM pillar. No global symmetry have still been firmly established.
However, it is highly probable that behind the quark and lepton flavor puzzle there exist underlying flavor symmetries. Another possibility is therefore that the DM stability is closely related to the flavor symmetry structures, for instance that the discrete symmetry at the origin of DM stability would be a remnant symmetry of a broken flavor group. This possibility has been recently considered  \cite{valle} on the basis of an $A_4$ global flavor symmetry which is dynamically broken to a
the $Z_2$ discrete group.

Finally we limited ourselves to the candidates whose relic density is determined by the thermal freeze-out of their annihilation. Of course there also exist DM candidates whose stability results from the fact that the allowed renormalizable couplings, which could induce a decay, would be very tiny. In particular if the only interaction causing DM decay is the gravitational coupling, one can have a stable enough DM candidate.
Similarly if one assumes very light right-handed neutrinos, of order KeV, and therefore tiny Yukawa couplings (not to induce too large neutrino masses) the lightest right-handed neutrino could be stable enough \cite{shapo}. Another DM candidate is of course the axion which if it has a very tiny mass, ${\cal O}(10^{-3}$~eV), can decay only to photons with naturally suppressed rates \cite{sikivie}. A KeV majoron is also an option \cite{valle2}.

\section{Conclusion}

The fact that the DM particle is stable on cosmological time scales is a peculiar property. It definitely needs an explanation.
On the basis of a gauge principle there are quite a few possible origins for this stability.
Along these mechanisms DM can be a scalar, a fermion, or even a gauge boson (or higher spin object such as the gravitino). The gauge symmetry invoked can be abelian or non abelian (both in their confined or unconfined phase).
Each mechanism leads to a characteristic phenomenology. In particular, 1) the existence of an exact gauge group results in extra radiation which might have many astrophysical effects, 2) accidental symmetry mechanisms are expected to lead to a rich indirect detection phenomenology from DM decay, such as intense $\gamma$ ray lines in the hidden vector DM setup, 3) high energy stabilization mechanism, based in particular on $SO(10)$, can have DM as the missing piece for unification of gauge couplings, 4) if no extra gauge group are assumed the DM must be part of a high $SU(2)_L$ multiplet with definite properties. Apart for the models where DM lies in the multi-TeV range, and for the models based on extra-radiation, all models can lead to specific signatures at the LHC.

\section*{Acknowledgements}
We thank X.~Chu for many discussions. Part of this work was realized in collaboration with C.~Arina, X.~Chu, M.~Frigerio, A.~Ibarra, M.~Tytgat and C.~Weniger. This work is supported by the FNRS-FRS, the IISN and the Belgian Science Policy (IAP VI-11).

\end{document}